\documentclass[12pt,preprint]{aastex}
\usepackage{amsmath}
\usepackage{graphicx}
\usepackage{natbib}

\newcommand{\skipthis}[1]{}

\def\um{$\mu$m}
\def\nh2d{$\rm{NH_2D}$}
\def\nh3{$\rm{NH_3}$}
\def\NH3{$\rm{NH_3}$}
\def\n2hp{$\rm{N_2H^+}$}

\def\msun{$M_\odot$}

\def\lsun{$L_\odot$}
\def\kms-1{km~s$^{-1}$}
\def\kms{km~s$^{-1}$}
\def\h2o{$\rm{H_2O}$}
\def\h2{$\rm{H_2}$}
\def\water{H$_2$O}
\def\meth{CH$_3$OH}

\def\cm2{$\rm{cm^{-2}}$}
\def\cm3{$\rm{cm^{-3}}$}
\def\dv{$\Delta V$}

\def\11{(1,1)}
\def\22{(2,2)}
\def\33{(3,3)}
\def\44{(4,4)}
\def\t21{$T_{21}$}
\def\r31{$R_{31}$}
\newcommand{\lsim}{${\raisebox{-.9ex}{$\stackrel{\textstyle<}{\sim}$}}$ }

\begin{document}

\slugcomment{Accepted to ApJL}

\shorttitle{Protostellar Outflow Induced Gas Heating}
\shortauthors{Wang et al.}

\title{Protostellar Outflow Heating in a Growing Massive Protocluster}
\author{
Ke Wang\altaffilmark{1,2},
Qizhou Zhang\altaffilmark{2},
Yuefang Wu\altaffilmark{1},
Hua-bai Li\altaffilmark{3},
and 
Huawei Zhang\altaffilmark{1}
}
\email{kwang@cfa.harvard.edu}
\altaffiltext{1}{Department of Astronomy, School of Physics, Peking University,
Beijing 100871, China}
\altaffiltext{2}{Harvard-Smithsonian Center for Astrophysics, 60 Garden Street,
Cambridge MA 02138, USA}
\altaffiltext{3}{Max-Planck Institute for Astronomy, K\"{o}nigstuhl 17, D-69117 Heidelberg, Germany}

\begin{abstract}
The dense molecular clump P1 in the infrared dark cloud (IRDC) complex G28.34+0.06
harbors a massive protostellar cluster at its extreme youth.
Our previous Submillimeter Array (SMA) observations revealed several
jet-like CO outflows emanating from the protostars, indicative of 
intense accretion and potential interaction with ambient natal materials.
Here we present the Expanded Very Large Array (EVLA) spectral line observations toward P1 in the
\nh3 (J,K) = \11, \22, \33 lines, as well as \water\ and class I \meth\ masers.
Multiple \nh3 transitions reveal the
heated gas widely spread in the 1 pc clump.
The temperature distribution is highly structured;
the heated gas is offset from the protostars, and
morphologically matches the outflows very well.
Hot spots of spatially compact, spectrally broad
\nh3 \33 emission features are also found coincident with the outflows.
A weak \nh3 \33 maser is discovered at the interface between an outflow jet and the ambient gas.
These findings suggest that protostellar heating may not be effective in suppressing
fragmentation during the formation of massive cores.
\end{abstract}


\section{Introduction}

The majority of stars, from low to high mass ones are born in clusters \citep{Lada2003}.
In a clustered environment, feedback from protostellar radiation and outflows can
play an important role in shaping their parental molecular cloud
and thereby regulate further cloud fragmentation and star formation.
In particular, protostellar feedback may be crucial for the formation of
massive stars which typically contain masses at least one order of
magnitude higher than the global Jeans mass in the molecular clump
($M_J \sim 1$ \msun\ for $n_{\rm H} \sim 10^5$ \cm3 and $T \sim 15$ K; 
\citealt{pillai06, rath06}).
Calculations have shown that
the accretion luminosity of a low mass protostar can heat gas to 100 K out to hundreds of AU, 
thereby raising the local thermal Jeans mass \citep{Masunaga2000, Krumholz2006}.
Subsequent numerical simulations indicate that protostellar heating from low
mass stars in the cluster may change the equation of state \citep{Larson1985}, and
suppress the fragmentation \citep{Krumholz2010b}, thereby aiding the formation of massive
cores. On the other hand, protostellar outflows in the cluster can also inject momentum and
energy into the local medium, and affect the evolution of the cloud \citep{Quillen2005,Curtis2010,WangP2010}.

To investigate the protostellar feedback,
we observed the dense molecular clump P1 in the 
IRDC complex G28.34+0.06.
Located in the southern part of G28.34+0.06,
P1 is one of the dense molecular clumps among the filamentary network in the IRDC complex
which stretches over 6 pc at a kinematic distance of 4.8 kpc 
\citep{Carey2000, rath06, wy08}.
P1 is a northeast-southwest (NE-SW) orientated molecular clump containing
a large mass reservoir of $10^3$ \msun\ within a size of 1 pc.
\cite{wy08} presented an \nh3 mosaic of the entire IRDC,
and found that the gas temperature decreases from 20 to 14 K
toward the inner part of P1, indicative of external heating.
A similar trend was found in the \nh3 line width,
suggesting a turbulence dissipation toward smaller scales.
Our SMA observations
resolved the P1 clump into a string of five cores (named SMA1--5)
containing 22--64 \msun\ in mass and $\lsim 0.1$ pc in size;
at higher angular resolution and better sensitivity,
the cores are further resolved into ten compact condensations (named, e.g. SMA2a--d)
with masses of 1.4--10.6 \msun\ and sizes of $\sim$0.01 pc,
indicative of a hierarchical fragmentation \citep{qz09,me11}.
In addition, all five cores are associated with
highly collimated CO (3--2) outflows.
The large mass reservoir and the ongoing accretion indicate
that P1 will eventually form a massive star cluster 
after the accretion is completed. 
This young protocluster provides an opportunity
to investigate how protostellar heating and outflows affect the parent molecular 
environment, and the formation of massive cores.
The new EVLA observations presented in this Letter reveal
the temperature profiles at unprecedented high angular resolution,
allowing us to evaluate the protostellar feedback in detail.

\section{Observations \label{sec:obs}}

The NRAO\footnote{The National Radio Astronomy Observatory
is a facility of the 
National Science Foundation operated under cooperative agreement by Associated Universities, Inc.}
EVLA was used to obtain the \nh3\ (J,K) = \11, \22, \33, 
as well as H$_2$O and CH$_3$OH maser line emission data
in 2010 during the EVLA early science phase. 
Table \ref{tab:obs} summarizes the observations.
The data were calibrated in CASA\footnote{\url{ http://casa.nrao.edu}}
and imaged using
CASA and Miriad\footnote{\url{ http://www.astro.umd.edu/$\sim$teuben/miriad,  http://www.cfa.harvard.edu/sma/miriad}}.

The \nh3 \11 and \22 observed in the EVLA C array
has a channel width of 0.2 \kms.
\cite{wy08} observed \nh3\ (1,1) and (2,2) toward the same phase center
in the VLA D array with a channel spacing of 0.6 \kms.
We combined the calibrated visibilities of \cite{wy08} with the new C array data for imaging.
To increase the signal-to-noise ratio in the final images,
we averaged the channel width to 0.3 \kms\ and
tapered the visibilities with a Gaussian function with a FWHM of 180 k$\lambda$.
We also made images with a 0.6 \kms\ channel width to 
compare with the images at 0.3 \kms\ resolution,
and found that the latter does not introduce additional features.
For the \nh3 \33 images we also averaged the channel width to 0.3 \kms\ 
to improve the signal-to-noise ratio.

\section{Results}
\subsection{Ammonia Emission \label{sec:nh3} }

Figure \ref{fig:mnt} presents moment 0, 1 and 2 maps of the 
\nh3 (1,1), (2,2) and (3,3) main hyperfine emission.
The NE-SW orientated filament is traced by 
the \nh3 (1,1) and (2,2) emission extremely well.
Ammonia cores coincide with all the SMA dust cores,
indicating that \nh3 traces the dense envelopes of the protostellar cores.
As illustrated in the moment 1 images of \nh3 (1,1) and (2,2),
there is a velocity gradient of 1.5 \kms\ over 1 pc along the filament.
This gradient is, however, not uniform; 
it is enhanced toward southwest starting at SMA5.
On the other hand, 
the velocity field from SMA1 through SMA4 is quite uniform, at about 79 \kms.
The moment 2 images show similar velocity dispersion\footnote{For a Gaussian line profile,
velocity dispersion $\sigma$ is related to FWHM line width \dv\ as
$\sigma = \Delta V / 2\sqrt[•]{2\mathrm{ln} 2}$.} 
of 0.4--0.8 \kms\
along the filament as traced by \nh3 (1,1) and (2,2).
The velocity dispersion is enhanced to several local maxima of 0.6--0.8 \kms\
which coincide with all the SMA dust cores except SMA3.
The enhancements may originate from a combination of spatially
unresolved motions such as rotation, infall, and outflow.

In addition to the main filament,
the \nh3 (1,1) emission also reveals a fainter northwest-southeast (NW-SE) orientated filament, 
crossing the main filament at SMA5,
making an $X$ shape.
This minor filament is located at a slightly different velocity, around 82 \kms.
Compared to the main filament, 
the minor filament shows weaker emission in \nh3 (1,1) and little emission in \nh3 (2,2), 
indicative of even colder gas.
The minor filament is evident in the previous \nh3 (1,1) image
at lower resolution \citep{wy06,qz09},
however, it does not correspond to any dust emission feature 
at 350--1300 \um\ \citep{rath06,rath10,qz09,chen10,me11}.
Therefore, the minor filament is remarkably less dense than the main filament.
Like the main filament,
the minor filament consists of a velocity gradient of 2 \kms\ over 1 pc;
it also hosts several \nh3 cores with sizes similar
to the \nh3 cores on the main filament.
It has a velocity dispersion of $\leq 0.4$ \kms\ in general,
but local enhancements of up to 0.5 \kms\ are also presented.
We will discuss the nature of the minor filament in Section \ref{sec:mag}.
In the following text, unless otherwise stated,
we refer the main NE-SW filament as ``the filament''.

It is interesting to note that the clumpy \nh3 (3,3) emission does not follow the dust filament at all.
We visually identified five representative peaks from the (3,3) moment 0 map,
labeled as A--E in Figure \ref{fig:mnt}.
None of these peaks coincide with any dust cores.
In fact, the only one that lies on the filament is peak D which
is between SMA4 and SMA5;
other peaks are offset from the filament.
Because \nh3 (3,3) is excited at an energy level of 125 K,
these ammonia peaks trace local temperature enhancements 
relative to the generally cold molecular clump.
Another important feature is the large velocity dispersion:
all the peaks show centrally condensed distribution of $\sigma > 0.8$ \kms,
and up to 2.2 \kms\ in peak B.
Peak B is located down stream of the red lobe of outflow SMA2a,
the most energetic outflow in P1 \citep{me11}.
The centroid velocity on peak B is redshifted relative to the filament,
consistent with the redshifted CO emission.
Indeed, all these \nh3 (3,3) peaks seem to trace the footprint of the
jet-like CO (3--2) outflows revealed by the SMA \citep{me11}.
We will discuss this further in Section \ref{sec:heat}.

\subsection{Maser Emission \label{sec:maser} }

\begin{figure}
\centering
\includegraphics[width=.64\textwidth,angle=0]{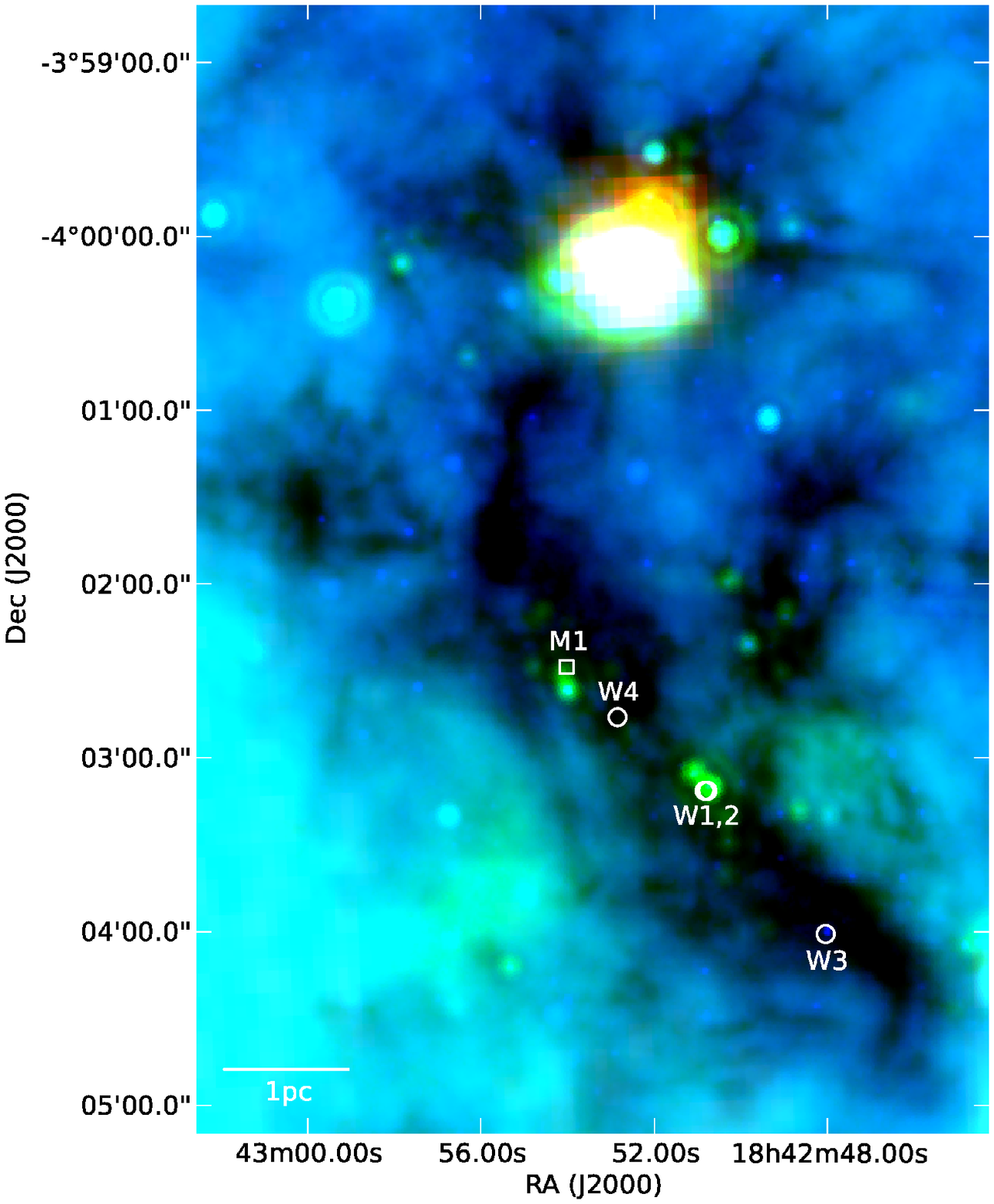}
\includegraphics[width=.32\textwidth,angle=0]{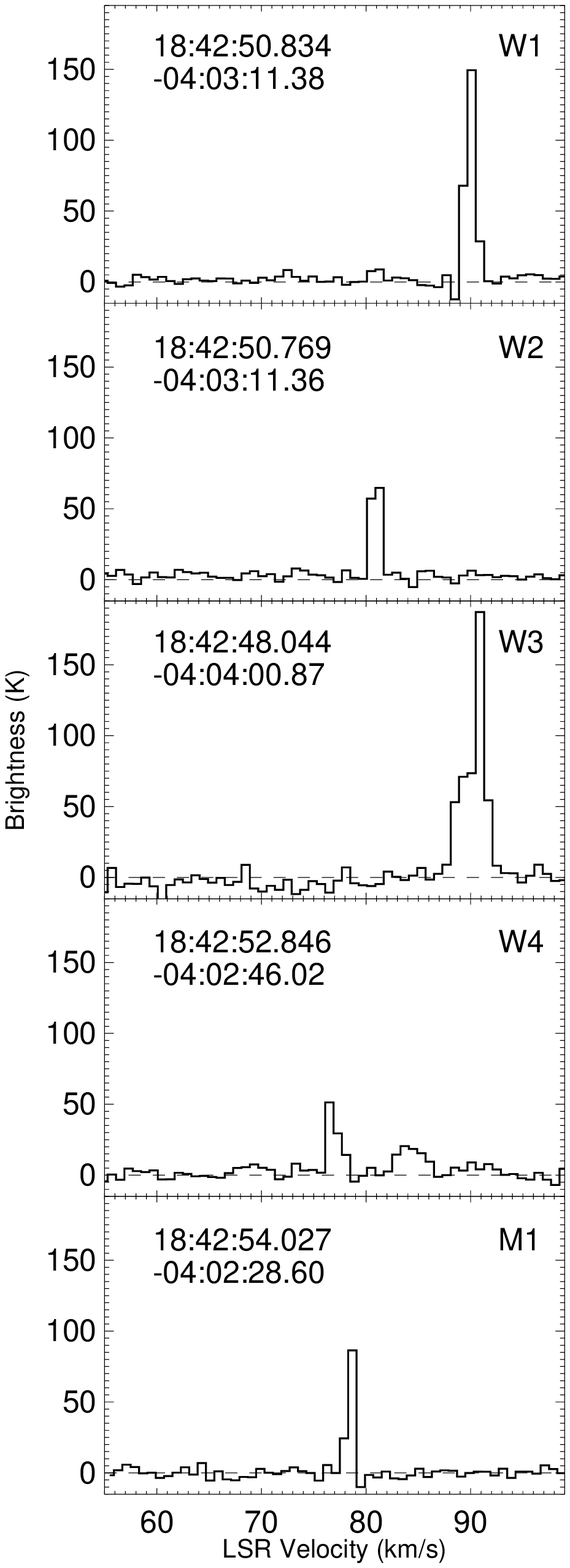}
\caption{
Detected water (W1--4) and methanol (M1) masers.
\textbf{Left:}
\emph{Spitzer} composite image (red/green/blue = 70/24/8 \um)
showing the IRDC G28.34+0.06 complex overlaid with the locations of the masers.
The \emph{Spitzer} data are taken from the GLIMPSE and MIPSGAL legacy projects \citep{GLIMPSE, MIPSGAL}.
\textbf{Right:}
Spectra of the masers.
Brightness temperatures are corrected for primary beam response.
}
\label{fig:maser}
\end{figure}

We report maser detections based on the following criteria:
(a) Peak flux $\geq$6$\sigma$;
(b) Emission continues at least two channels with flux $\geq$4$\sigma$; and
(c) Within the primary beam response of $\geq$0.3.
We adopt these criteria because at some channels the images are contaminated 
by the side lobes of a strong maser located at P2, well outside of the primary beam.
Using the criteria, 
we identify 4 water masers (W1--4) and 1 methanol maser (M1),
as presented in Figure \ref{fig:maser}.
Positions of the masers are obtained by fitting a two dimensional Gaussian
to the brightness distribution of the maser emission.
The positional uncertainty is $\leq$0$''.02$.
All masers are either coincident with or close to infrared point sources at 8--24 \um.

Comparing the maser positions and velocities with previous observations
made with VLA (\citealt{wy06}, resolution $\sim$2$''$) 
and GBT (\citealt{chambers09}, resolution $\sim$30$''$),
we find that W1, W3, and W4 are new features,
while W2 and M1 are spatially and spectrally coincident with two masers reported before.
Both  W2 and M1 have
been varying in peak flux during the last several years:
W2 was 2.4 Jy on 2005 September 23/24,
then brightened to 6 Jy on 2006 December 5,
and dimmed dramatically to 0.03 Jy on 2010 November 24;
M1 dimmed from 0.24 Jy on 2006 December 5 to 0.02 Jy on 2010 November 24.
On the other hand,
W1 and W3 were either newly emerged
or experienced brightening between 2006 and 2010,
otherwise they should have been detected by \cite{chambers09}.
W4 is below the detection limit of \cite{chambers09}.

Among all the five detected masers, only two (W1, W2) 
are associated with \nh3 emission above $6\sigma$ significance.
W1 and W2 locate right on top of the dust condensations
SMA2a and SMA2b, respectively,
and thus are presumably excited by the protostars embedded 
in the same dust condensations.
The position discrepancy between the \water\ maser with its exciting protostar
is within a fraction of 1000 AU.

\section{Discussion} \label{sec:discuss}

\subsection{Outflow Heating} \label{sec:heat}

\begin{figure}
\centering
\includegraphics[width=0.62\textwidth,angle=0]{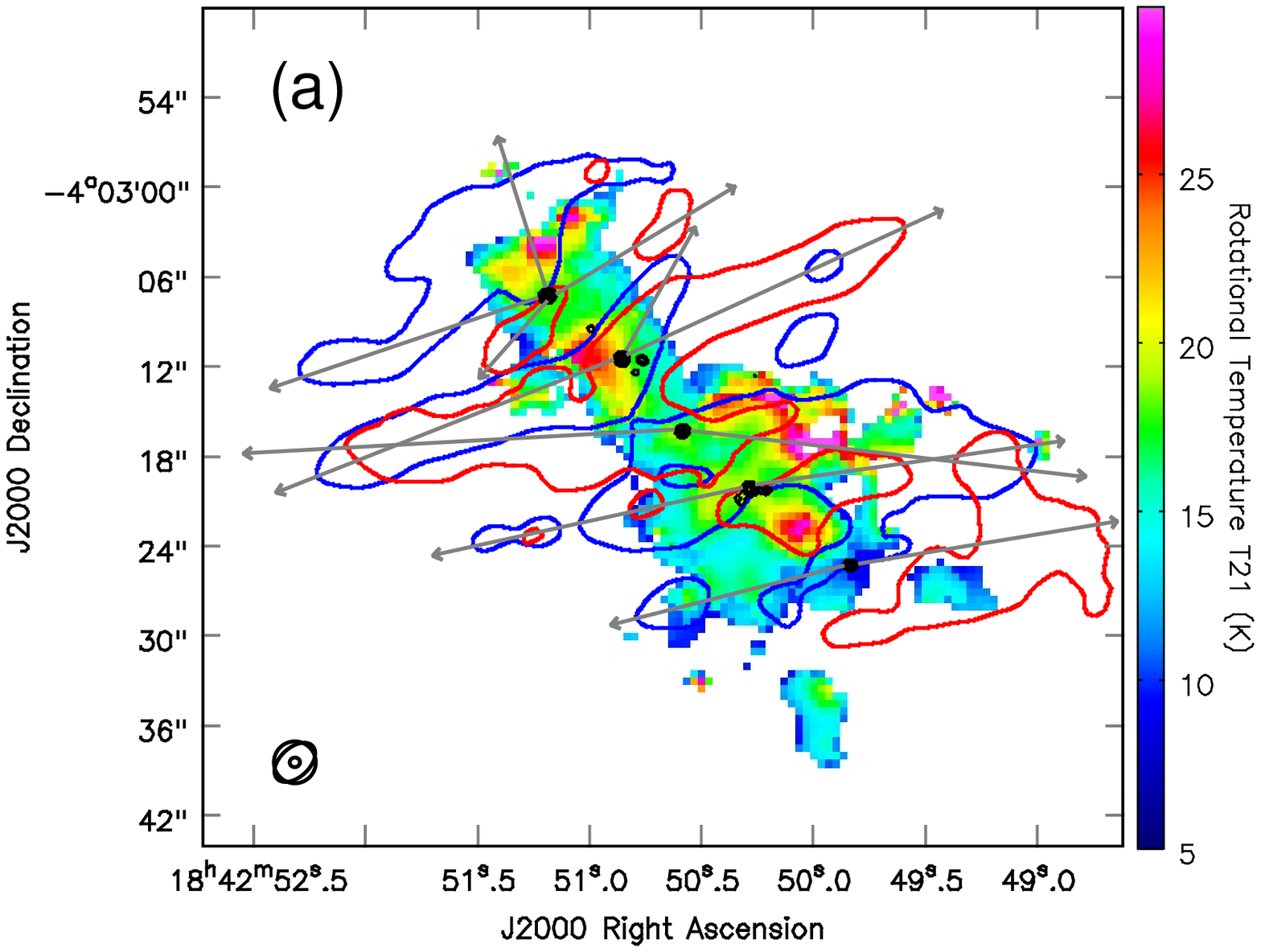}
\includegraphics[width=0.6\textwidth,angle=0]{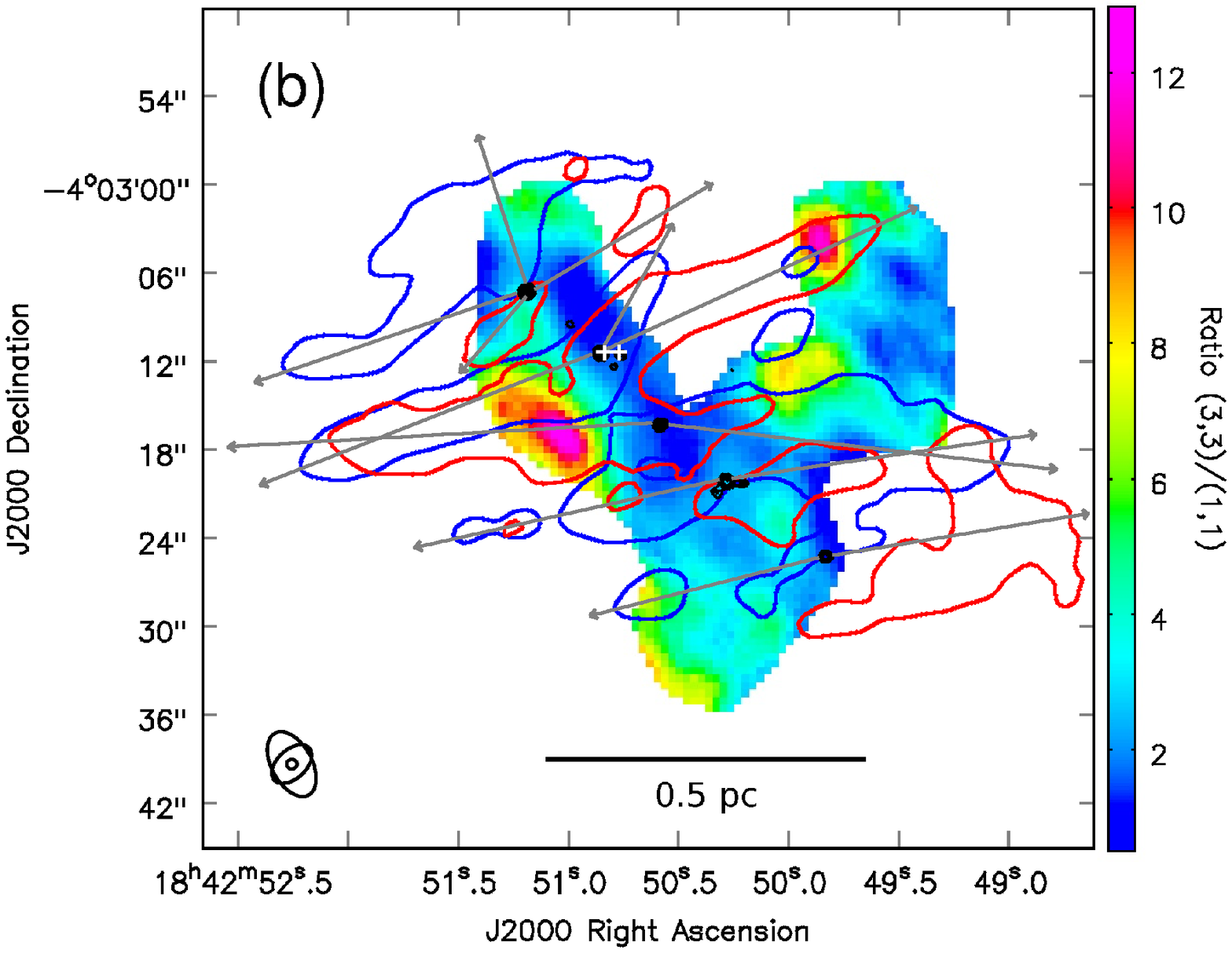}
\caption{
Morphological match between heated gas and protostellar outflows.
Plotted in color scales are 
\textbf{(a)} \t21,
the rotational temperature derived from \nh3 \11\ and \22, and 
\textbf{(b)} \r31,
the integrated intensity ratio of \nh3\ (3,3)/(1,1).
The arrows and blue/red contours outline the blue-/redshifted CO\,(3--2) outflows,
and the compact black contours represent the five groups of dust condensations,
SMA1--5, from upper-left to bottom-right \citep{me11}.
The two white crosses on top of SMA2a and SMA2b
represent water masers W1 and W2, respectively.
On the bottom-left corners are the beams for 
dust continuum, CO\,(3--2), and \t21 (\r31),
from the smallest to the largest respectively.
Note that masks of $5\sigma$ were applied to both maps, 
and additional manual clipping was also performed to \r31\
in order to avoid noisy edge pixels.
}
\label{fig:heat}
\end{figure}

Intensity ratios of the main and hyperfine components from multiple \nh3 (J,K)
transitions constrain rotational temperature 
which well approximates {kinetic} temperature in the regime of $<$20 K
\citep{Ho1983, Walmsley1983}.
In the following analysis, we assume that the rotational temperature is the same as the kinetic temperature when the former is $<$20 K.
Following the procedure of \cite{Ho1983},
we compute a rotational temperature (\t21) map based on the
\nh3 \11 and \22 images.
The \nh3 images were first smoothed to a common beam of $2''.8$
before computing the \t21 map.
The uncertainty of \t21 is estimated to be 3 K.
Figure \ref{fig:heat}(a) shows \t21 with
the dust condensations and the CO (3--2) outflows superposed.
The \t21 map reveals a highly structured temperature profile ranging from  8 to 30 K,
opposite to the smooth temperature profiles observed in other IRDCs \citep{Ragan11}.
This may be due to beam dilution since the data in \cite{Ragan11} have a larger beam.
A number of remarkable enhancements ($T_{21}>26$ K)
are present among the entire map, but surprisingly
none of the temperature enhancements coincide with any dust condensations.
Instead, all the \t21 enhancements spatially match the footprint of the outflows.

\begin{figure}
\centering
\includegraphics[width=.4\textwidth,angle=0]{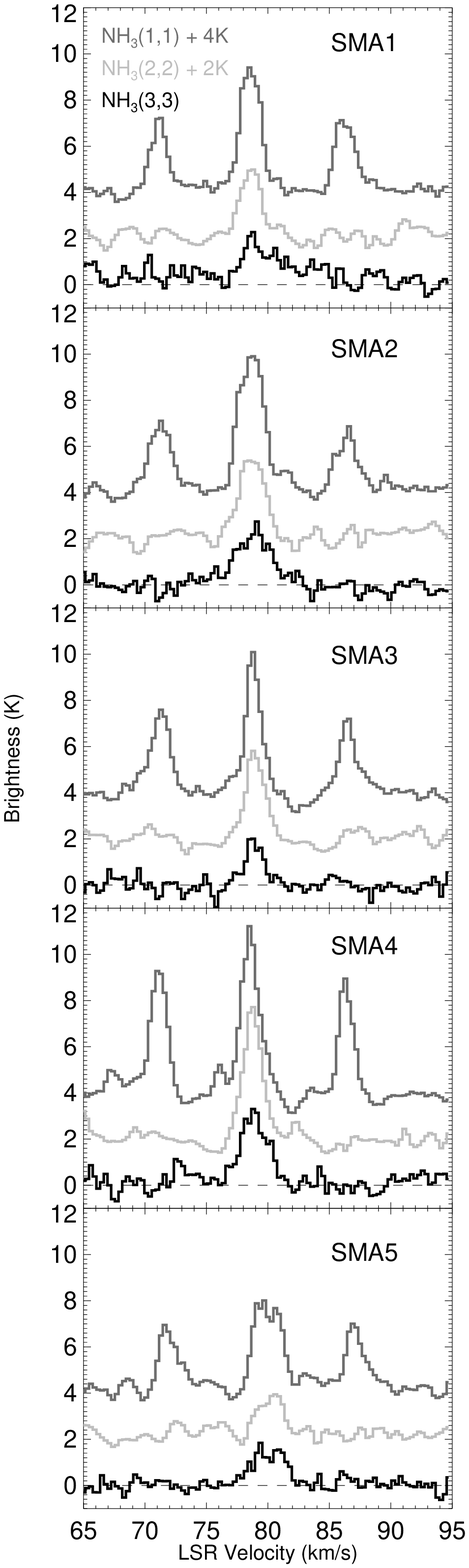}
\includegraphics[width=.4\textwidth,angle=0]{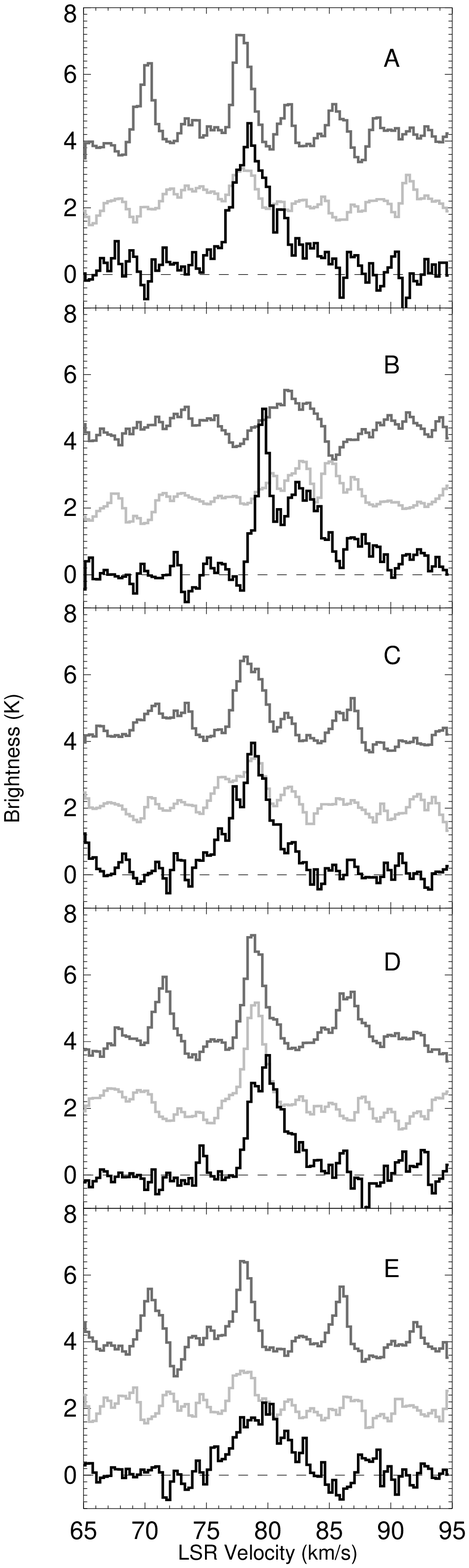}
\caption{
EVLA \nh3\ spectra extracted from SMA dust peaks (SMA1--5) 
and \nh3\ (3,3) peaks (A--D).
Note at peak B a spectroscopically unresolved \nh3 \33 maser at about 84 \kms.
\textit{(Panel E was truncated in formal publication due to the space limit of ApJ Letters.)}}
\label{fig:spec}
\end{figure}

Since the \nh3 \33 is more widely spread than the \11 toward the northwest
where the signal-to-noise ratio of \11 prohibits us from computing opacity,
it is difficult to derive a reliable rotational temperature map from the \nh3 \11 and \33 lines.
Instead, we compute the intensity ratio of the main hyperfine of \33/\11, \r31,
as plotted in Figure \ref{fig:heat}(b).
To compute the intensity ratio, the \nh3 \11 image was first smoothed to the beam of the \33 image.
The intensity ratio is positively correlated with the rotational temperature \citep{Ho1983, qz02}.
For reference, 
assuming optically thin emission and the same abundance for the ortho and para \nh3 
\citep{qz07},
$R_{31}=0.5$, 2, and 5 correspond to rotational temperatures 
of $T_{31} = 31$, 55, and 109 K, respectively.
Like the \t21 map, the \r31 map is also highly structured.
\r31 varies continuously from 0.3 to 12.5 over the map.
Again several hot spots represented by \r31 peaks 
(corresponding to \nh3 \33 emission peaks A--E)
match the outflow footprint and are offset from the dust condensations.
In addition to the peaks,
there is a low \r31 valley along SMA2 to SMA3.
Optically thin thermal \nh3 \33 emission should have \r31 $\leqslant 6.4$ \citep{Ho1983}.
Higher \r31 values are seen toward peaks A--C,
indicative of non-thermal emission.
Toward peak B, the (3,3) emission is weakly amplified (Figure \ref{fig:spec}),
resembling an \nh3 \33 maser similar to those found in 
other massive star formation regions where outflows are present
\citep[e.g.][]{Mangum1994, qz95, Beuther2007, Brogan2011}.
However the spatial resolution is insufficient to constrain the brightness of this emission
feature.

The morphological match between the heated gas and the outflow footprint suggests that
the heating in the P1 region is due to protostellar outflows.
Indeed, all the hot spots exclusively coincide with large \nh3 \33 line widths
compared to relatively small line width toward the dust condensations (Figure \ref{fig:spec}),
likely indicative of gas motion and extra turbulence injected by the outflow shocks.
The outflow heating has two important features.
First, the heating alters thermal property of the gas 
throughout the clump, from a distance 0.1 pc up to 0.5 pc away from the driving protostars.
Second, the heating is localized and far from well mixed.
The temperature at the low \r31 valley is not high enough
to suppress further fragmentation into Jeans mass of 1 \msun\ \citep{me11}.
We probably have witnessed through EVLA and SMA the 
very beginning of an outflow induced gas heating.

\subsection{Radiative Heating}

It is apparent that protostellar outflows can introduce significant heating
to the ambient gas.
What is the effect of protostellar radiation? Is radiative heating from low mass
protostars sufficient to suppress fragmentation and aid the formation of massive cores?
The fact that all dense cores correspond with low \r31 is intriguing.
In addition, the temperature map in Figure \ref{fig:heat}(a) indicates that dense cores have typical
temperatures of 18 K, and are offset from the high temperature peaks. The temperature
distributions suggest that the heating of low mass protostars may not play a
significant role in the formation of dense cores detected with the SMA.
Otherwise, massive cores should be found at the peaks of the temperature distribution.
\cite{Longmore2011} made a similar statement for a hot molecular core G8.68--0.37.

It is also interesting to note that the protostars embedded in the SMA cores
have not produced significant heating.
SMA1 to SMA4 have a rather uniform \t21 of 18 K,
while SMA5 has a lower \t21 of 9 K.
We caution that the \t21 map has a $2''.8$ beam,
so any temperature structure has been smoothed to a spatial scale of 0.065 pc.
How much luminosity is required to heat the gas to 18 or 9 K 
at a radius of 0.065 pc from the central source?
Assuming thermal equilibrium between gas and dust,
the luminosity can be estimated following the equation \citep{scoville76,garay99}:
$$
T_\mathrm{dust} = 65
\left( \frac{0.1\, \mathrm{pc}}{r} \right) ^{2/(4+\beta)}
\left( \frac{L_\mathrm{star}}{10^5 L_\odot} \right) ^{1/(4+\beta)}
\left( \frac{0.1}{f} \right) ^{1/(4+\beta)} \,,
$$
where $\beta = 1.5$ is the dust emissivity at far infrared wavelengths \citep{rath10},
$f = 0.08 \, \rm{cm}^2 \, \rm{g}^{-1}$ is the dust emissivity at 50 \um\ \citep{scoville76},
and $r = 0.065$ pc is the radius from the central protostar.
We estimate a luminosity of 30 \lsun\ for each protostar embedded in SMA1, SMA2a, SMA3, and SMA4a, 
and 1 \lsun\ for the protostar SMA5.
These values should be taken with caution because 
the beam dilution may introduce large uncertainty in both temperature and radius.
Nevertheless, we note that the luminosity of SMA2a is in rough agreement 
with the luminosity inferred from SED for the 24 \um\ source
in the SMA2 vicinity \citep[$\sim$10$^2$ \lsun,][]{wy08}.
A lack of significant heating from the growing protostars is consistent with non detection
of line emission from organic molecules \citep{qz09,me11}.

\subsection{Origin of the Minor Filament} \label{sec:mag}

\begin{figure}
\centering
\includegraphics[height=\textwidth,angle=-90]{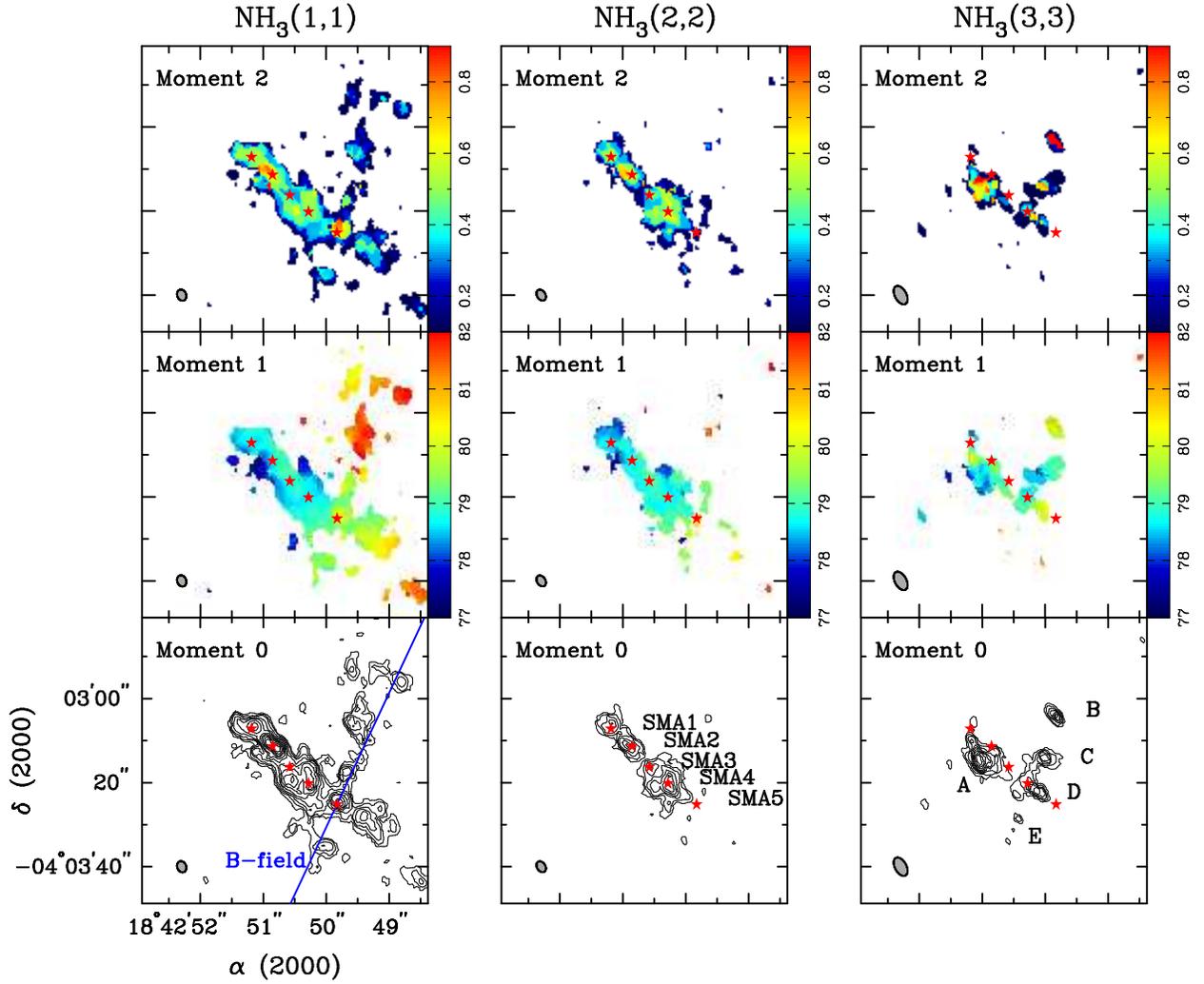}
\caption{
Moment maps of the main hyperfine emission of \nh3 (1,1), (2,2) and (3,3),
shown on the left, middle and right columns, respectively.
The contours are plotted in steps of $6\sigma$, which is
0.5 mJy beam$^{-1}$ \kms\ for (1,1) and (2,2),
while 1 mJy beam$^{-1}$ \kms\ for (3,3).
Units of the color bars are \kms.
The stars mark the five dust cores (SMA1--5) which are further resolved 
into smaller condensations (the black contours in Figure \ref{fig:heat}).
The blue line on the bottom-left panel indicates the direction of the 
mean magnetic field (Section \ref{sec:mag}).
The filled ellipses on the bottom-left corners represent synthesized beams.
}
\label{fig:mnt}
\end{figure}

The origin of the low-density minor filament is rather interesting.
A low-density streak perpendicular to a high-density oblate region
is expected if magnetic field is dynamically important compared to self-gravity and turbulence
during the formation of the system \citep{hb11}.
In this case,
gravitational contraction along the field lines is more efficient,
resulting in a high-density region perpendicular to the magnetic field,
while on the other hand turbulence is also channeled by the field and gives rise to
streaks aligned with the magnetic field in the lower-density region.
This phenomenon has been seen in simulations \citep{Price2008,Nakamura2008}
and in observations of NGC 2024 in Orion \citep{hb11}.
In G28.34-P1, the $X$-shaped filamentary system projected on the sky is almost perpendicular.
If the system were formed from the same mechanism,
one should expect a mean magnetic field 
along the minor filament and perpendicular to the main filament.
Based on the JCMT dust polarization data cataloged by \cite{Matthews2009},
we derived a mean B-field direction of $\mathrm{P.A.} = -25^{\circ}.6 \pm 45^{\circ}$
as labeled on Figure \ref{fig:mnt}.
Because the main filament has a P.A. of $48^{\circ}$ \citep{me11},
the separation angle between the mean B-field and the main filament is $73^{\circ}.6$.
The mean magnetic field at the clump scale is indeed along with the minor filament
and roughly perpendicular to the main filament.
The large deviation in the B-field direction
may indicate that the strong outflows have started to disturb
the initially ordered field lines in the inner part of P1 \citep{hb09}.
Therefore we speculate that the minor filament originates from turbulence anisotropy
during the formation of the $X$-shaped filamentary system,
and that the formation process has been governed by 
the interplay between strong magnetic field, self-gravity, and turbulence.

\acknowledgements
We thank Edward T. Chambers for providing the spectra of the masers detected in G28.34+0.06.
K.W. acknowledges the support from the SMA predoctoral fellowship and the China Scholarship Council.
Q.Z. acknowledges the support from the Smithsonian Institution Endowment Funds.
This research was supported in part by the NSFC grants 10733030, 10873019 and 11073003.

{\it Facilities:} \facility{EVLA}

\begin{deluxetable}{lcccc}
\tablecolumns{5}
\tablewidth{0pc}
\tabletypesize{\footnotesize}
\tablecaption{Observational Parameters \label{tab:obs}}
\tablehead{
\colhead{Parameter/Line} & \colhead{\nh3 \11} & \colhead{\nh3 \22} &
\colhead{\nh3 \33} & \colhead{\water/\meth}
}
\startdata
Observing date 	&2010Oct09	&2010Oct11	&2010May09	&2010Nov24	\\
Configuration	&C	&C	&D	&C	\\
Rest frequency (GHz)	&23.694	&23.722	&23.870	&22.235/24.959	\\
Primary beam ($'$)	&2	&2	&2	&2	\\
Gain calibrator	&J1851+005	&J1851+005	&J1851+005	&J1851+005	\\
Flux calibrator \tablenotemark{a}	&3C48	&3C48	&3C48	&3C48	\\
Bandpass calibrator	&3C454.3	&3C454.3	&3C454.3	&3C454.3	\\
Phase center \tablenotemark{b}	&I	&I	&II	&II	\\
Integration time (minute)	&95	&95	&21	&15/15	\\
Bandwidth (MHz)	&4	&4	&4	&4	\\
Polarization &dual	&dual	&dual	&full	\\
Naive channel width (\kms)	&0.2	&0.2	&0.2	&0.8	\\
Final channel width (\kms)	&0.3	&0.3	&0.3	&0.8	\\
Synthesized beam ($''$) \tablenotemark{c}	&$2.7 \times 2.2$	&$2.8 \times 2.0$	&$4.8 \times 2.7$	&$1.3 \times 1.0$/$1.5 \times 0.9$	\\
RMS noise (mJy/beam) \tablenotemark{d}	&1.2--1.5	&1.2--1.5	&3	&1.2--1.5/0.8	\\
Convention factor (K/Jy)	&367	&388	&165	&1939/1484	\\
Weighting method	& natural & natural & natural & robust/natural \\
\enddata
\tablenotetext{a}{Absolute flux is accurate to about 5\%.}
\tablenotetext{b}{Phase center I at (R.A., decl.)$_{\mathrm J2000}=
18^{\mathrm h}42^{\mathrm m}50^{\mathrm s}, -04^{\circ}03'30''$;
II at 
$18^{\mathrm h}42^{\mathrm m}50^{\mathrm s}.82, -04^{\circ}03'11''.3$.}
\tablenotetext{c}{
Final \nh3 \11 and \22 images are made by adding archival D array data (Section \ref{sec:obs}).
}
\tablenotetext{d}{$1\sigma$ rms noise measured in final images.}
\end{deluxetable}

\end{document}